\documentclass[11pt,aps,superscriptaddress]{revtex4}

\usepackage{graphicx}
\usepackage{dcolumn}
\usepackage{bm}

\begin{document}                
\title{Quantal Density Functional Theory of the Hydrogen Molecule  }

\author{Xiao-Yin Pan and Viraht Sahni}
\affiliation{Department of Physics, Brooklyn College and The
Graduate School of the City University of New York,365 Fifth Avenue, New York, New
York 10016. }
%

\begin{abstract}                
In this paper we perform a Quantal Density Functional Theory (Q-DFT) study of the Hydrogen molecule
 in its ground state.  In common with traditional Kohn-Sham density functional theory (KS-DFT),
  Q-DFT transforms the interacting system as described by Schrodinger theory, to one of noninteracting
  fermions -- the S system -- such that the equivalent density, total energy, and ionization potential are obtained.
   The Q-DFT description of the S system is in terms of  `classical' fields and their quantal sources
    that are quantum-mechanical expectations of Hermitian operators taken with respect to the
     interacting and S system wave functions..  The sources, and hence the fields, are separately
      representative of all the many-body effects  the S system must account for, viz.
      electron correlations due to the Pauli exclusion principle, Coulomb repulsion and
      Correlation-Kinetic effects.  The local electron-interaction potential energy of each model
      fermion is the work done to move it in the force of a conservative effective field that is the
      sum of the individual fields.  The Hartree, Pauli, Coulomb, and Correlation-Kinetic  energy
       components of the total energy are also expressed in virial form  in terms of the corresponding fields.
        The highest occupied eigenvalue of the S system is the negative of the ionization potential energy.
         The Q-DFT analysis  of the Hydrogen molecule is performed employing the highly accurate correlated wave
         function of Kolos and Roothaan.  The structure of the sources -- the density, Fermi-Coulomb,
         Fermi, and Coulomb holes -- as a function of the electron position are obtained, and
         from them the corresponding fields. (To our knowledge, these are the first accurate graphs of the
         Fermi-Coulomb and Coulomb holes as a function of electron position for the Hydrogen molecule.) As a consequence of the symmetry of the molecule,
          the individual fields  -- Hartree, Pauli, Coulomb, Correlation-Kinetic -- are all \emph{antisymmetric}
          about the center of the nuclear bond.
          Thus, the electron-interaction potential energy, and its Hartree, Pauli,
          Coulomb, and Correlation-Kinetic components are each \emph{symmetric} about this center.
          The Coulomb correlation and Correlation-Kinetic fields, and hence their contributions to the
          potential and total energy are an order of magnitude smaller than those due to the Hartree and Pauli terms.
           However, the Correlation-Kinetic contribution is more significant than that due to Coulomb correlations.
            This new fact is important to the construction of approximate KS-DFT `correlation' energy functionals for
             molecules.  Finally, there is a striking similarity in the structure of the
              various sources, fields, and potential energies of the Hydrogen molecule for
              electron positions in the positive half-space encompassing one nucleus, and those of the Helium atom.

\end{abstract}
\maketitle
\newpage
\section{ Introduction}
In this paper we analyze the Hydrogen molecule ($H_{2}$) in its ground-state electronic
 configuration $(\sigma_{g}1s)^{2}$ from the perspective of time-independent Quantal density
 functional theory (Q-DFT) \cite{1,2,3,4,5,6,7,8}.
The \textit{in principle exact} framework of Q-DFT for ground and
excited states, both nondegenerate and degenerate, has been
demonstrated by application to exactly solvable model atomic
systems \cite{3, 5,6,7,8} as well as by the use of essentially
exact atomic correlated wave functions \cite{2,4,9}.
 In its approximate form, Q-DFT has been applied to atoms, atomic ions, atoms
 in excited states,
 and positron binding, as well as to the many-electron inhomogeneity at metallic
 surfaces and metallic clusters.
 We refer the reader to the review articles of Refs \cite{2,10} for further
  references on these applications.
 This paper constitutes a first step in the application of Q-DFT to molecules.
 Here we present the essentially exact analysis of the $H_{2}$ molecule via Q-DFT by
  employing the highly
 accurate correlated wave function of Kolos and Roothaan \cite{11}.
Beyond the understandings achieved, a principal attribute of the
calculation is the knowledge that the structure of the
corresponding Q-DFT properties for other diatomic molecules will
then be qualitatively similar. Furthermore, these essentially
exact  properties can be used as the basis for comparison and
testing of various approximations within Q-DFT prior to their
application to
more complex molecules.\\

 Q-DFT, in common with traditional Kohn-Sham density functional theory (KS-DFT)\cite{12},
 maps a system of electrons in an external field ${\bf F}^{ext} = -\nabla v({\bf r})$ in their
  \textit{ground} state to one of \textit{noninteracting} fermions in their \textit{ground} state with
  equivalent density.  The equivalent ground-state energy and ionization potential are thereby also obtained.
  The model system of noninteracting fermions is referred to as the S system, S being a mnemonic for a single
   Slater determinant. (Within the framework of Q-DFT, it is also possible \textit{in principle} to map
   into an S system in which the noninteracting fermions are in an \textit{excited} state.)  The local
   (multiplicative) effective potential energy $v_{s}({\bf r})$ of the model fermions is the sum of the
   external $v({\bf r})$ and an electron-interaction $v_{ee}({\bf r})$ potential energy, the latter
    being representative of all the electron correlations the S system must account for.
    These correlations are those due to the Pauli exclusion principle, Coulomb repulsion,
    and Correlation-Kinetic effects.  The Correlation-Kinetic contribution is a consequence of
    the difference in the kinetic energies of the interacting and noninteracting systems with equivalent density.
     In Q-DFT, the potential energy $v_{ee}({\bf r})$ is defined as the work done to move a model fermion
     in the force of a conservative `classical' field. The components of this field each separately represent
      a different electron correlation. The sources of these component fields are quantal in that they are
       expectations of Hermitian operators taken with respect to the Schr{\" o}dinger and S system wave functions.
       The Pauli (exchange), Coulomb (correlation), and Correlation-Kinetic components of the total energy are
        also separately expressed in integral virial form in terms of the fields representative of these correlations.
        The highest occupied eigenvalue of the S system differential equation is the negative of the ionization
        potential\cite{13}.
\\

The traditional KS-DFT description of the S system differs from
that of Q-DFT in the following manner.  Traditional theory   is in
terms of the ground state energy functional $E[\rho]$ of the
density $\rho({\bf r})$, and of its functional derivative. In
KS-DFT, the many-body correlations noted above are embedded in the
KS
electron-interaction energy functional $E^{KS}_{ee}[\rho]$.  The corresponding
electron-interaction potential energy of the noninteracting fermions is defined as the
 functional derivative of this functional taken at the true ground state density value:
 $v_{ee}({\bf r}) = \delta E^{KS}_{ee}[\rho]/\delta \rho({\bf r})$. Within KS-DFT, it is common practice to
 subtract the known Hartree or Coulomb self energy functional $E_{H}[\rho]$ from $E_{ee}[\rho]$ , thereby defining the
  KS `exchange-correlation' energy functional $E^{KS}_{xc}[\rho]$ and its functional derivative
  $v_{xc}({\bf r}) = \delta E^{KS}_{xc}[\rho]/\delta \rho({\bf r})$.  The functionals
  ($E^{KS}_{ee}[\rho], E^{KS}_{xc}[\rho]$) and their respective derivatives
   ($v_{ee}({\bf r}), v_{xc}({\bf r})$) are therefore also representative of the Pauli
   and Coulomb correlations and Correlation-Kinetic effects. KS-DFT, however does not describe how
    the different electron correlations are incorporated in the functionals ($E^{KS}_{ee}[\rho], E^{KS}_{xc}[\rho]$)
    and hence how they are represented in their functional derivatives.
     Furthermore, the functionals ($E^{KS}_{ee}[\rho], E^{KS}_{xc}[\rho]$) are themselves unknown.
      As such, even if the exact wave function of an interacting system were known, it is not possible to
      construct the corresponding S system directly by following the prescription of KS-DFT.
      Hence, the potential energy $v_{xc}({\bf r})$ is usually constructed indirectly via
      density-based methods [14-17] that employ knowledge of the `exact' density as determined
      from \textit{ab initio} calculations.\\

In Section II we give a brief description of ground state Q-DFT.  Section III is a
description of the various quantal sources, fields, energies and potential
energies pertaining to the S system as determined via Q-DFT employing the
$51$-parameter correlated wave function of Kolos-Roothaan. Concluding remarks
are made in Section IV. \\

\section{ Q-DFT OF A NONDEGENERATE GROUND STATE}
The Schr{\"o}dinger equation for a system of $N$ electrons in an
external field $\textbf{F}^{ext}({\bf r}) = -\bm{\nabla} v({\bf
r})$, and in a
 nondegenerate ground
 state,  is
\begin{equation}
[\hat{T}+\hat{V}+\hat{U}]\Psi({\bf X})=E \Psi ({\bf X}),
\end{equation}
where $\hat{T} = -\frac{1}{2} \sum_{i} \nabla_{i}^{2},\;\;\;
\hat{V}= \sum_{i} v({\bf r}_{i})$, and $\hat{U} = \frac{1}{2}
\sum^{\prime}_{i,j} \frac{1}{|{\bf r}_{i}-{\bf r}_{j}|}$ are the
kinetic energy, local external potential energy, and
electron-interaction potential energy operators, $\Psi({\bf X})$
and $E$ are the ground  state wave function and energy, with $
{\bf X}={\bf x}_{1}, {\bf x}_{2}, \dots,{\bf x}_{N} $, ${\bf
x}={\bf r}\sigma$, and ${\bf r}$ and  $\sigma$ the spatial and
spin coordinates. The ground state electronic density
 is the expectation
\begin{equation}
\rho({\bf r}) = \left\langle \Psi|\hat{\rho}|\Psi\right\rangle,
\end{equation}
where $\hat{\rho} = \sum_{i}\delta(\bf{r}-{\bf r}_{i})$ is the
Hermitian density operator. The corresponding spinless single
particle density matrix is the expectation
\begin{equation}
\gamma({\bf r}{\bf r}^{\prime})= \left\langle \Psi|\hat{ \gamma
}|\Psi\right\rangle,
\end{equation}
where the Hermitian operator $\hat{\gamma} = \hat{A}+i\hat{B}$,
$\hat{A}= \frac{1}{2}\sum_{j} [\delta({\bf r}_{j}-{\bf
r})T_{j}({\bf a})+ \delta({\bf r}_{j}-{\bf r}^{\prime})
T_{j}(-{\bf a})]$, $\hat{B}=-\frac{i}{2}\sum_{j} [\delta({\bf
r}_{j}-{\bf r})T_{j}({\bf a})- \delta({\bf r}_{j}-{\bf
r}^{\prime}) T_{j}(-{\bf a})]$, $T_{j}({\bf a})$ is a translation
operator, and ${\bf a}={\bf r}^{\prime}- {\bf r}$. The diagonal
matrix element of $\gamma({\bf r}{\bf r}^{\prime})$ is the
density: $\gamma({\bf r}{\bf r})=\rho({\bf r})$. The ground state
energy is the expectation
\begin{equation}
E=\left\langle \Psi|\hat{ T }+\hat{V}+\hat{U}|\Psi\right\rangle =T+E_{ext}+E_{ee},
\end{equation}
with the kinetic energy $T=\left\langle
\Psi|\hat{T}|\Psi\right\rangle$, the external potential energy
$E_{ext}=\left\langle \Psi|\hat{V}|\Psi\right\rangle=\int
\rho({\bf r})v({\bf r}) d {\bf r}$, and the electron-interaction
energy $E_{ee}=\left\langle \Psi|\hat{U}|\Psi\right\rangle$. The
ionization potential  is $I=E^{ion}-E$, where $E^{ion}$ is the
energy of the resulting ion when the
least bound electron is removed to infinity.\\

The  differential equation for the  S-system in its ground state
that leads to the  same density $\rho({\bf r})$ as that of the
electrons is
\begin{equation}
[-\frac{1}{2}\nabla^{2}+v({\bf r}) +
v_{ee}({\bf r})]\phi_{i}({\bf x})=\epsilon_{i}\phi_{i}
({\bf x}); \;\;\;\; i=1,\dots,N,
\end{equation}
where $v_{ee}({\bf r})$ is the  electron-interaction potential
energy of the noninteracting fermions. The S system wave function
is the Slater determinant $\Phi\{\phi_{i}({\bf x})\}$ of the
orbitals $\phi_{i}({\bf x})$, so that the density is the
expectation
\begin{equation}
\rho({\bf r})=\left\langle \Phi \{\phi_{i}\}|
\hat{\rho}|\Phi\{\phi_{i}\} \right\rangle =\sum_{i} \sum_{\sigma}
|\phi_{i}({\bf r}\sigma)|^{2},
\end{equation}
and the corresponding spinless Dirac density matrix
is the expectation
\begin{equation}
\gamma_{s}({\bf r}{\bf r}^{\prime})= \left\langle \Phi
\{\phi_{i}\}| \hat{{\bf \gamma}}|\Phi\{\phi_{i}\} \right\rangle
=\sum_{i} \sum_{\sigma} \phi_{i}^{*}({\bf r}\sigma)\phi_{i}({\bf
r}^{\prime}\sigma).
\end{equation}
\\

The potential energy $v_{ee}({\bf r})$ is the work done to move
the model fermion from a reference point at infinity to its
position at ${\bf r}$ in the force of a \textit{conservative}
effective field $\bm{\mathcal{F}}^{e\!f\!f}({\bf r})$:
\begin{equation}
v_{ee}({\bf r}) = - \int_{\infty}^{{\bf r}}
\bm{\mathcal{F}}_{k}^{e\!f\!f}({\bf r}^{\prime})\cdot d{\bf l}^{\prime}.
\end{equation}
The field $\bm{\mathcal{F}}_{k}^{e\!f\!f}({\bf r})$ is the sum
of an electron-interaction field $\bm{\mathcal{E}}_{ee}({\bf r})$
representative of Pauli and Coulomb correlations, and a
Correlation-Kinetic field $\bm{\mathcal{Z}}_{t_{c}}({\bf r})$ that is representative of the
correlation contribution to the kinetic energy:
\begin{equation}
\bm{\mathcal{F}}^{eff}({\bf r})=
\bm{\mathcal{E}}_{ee}({\bf r})+\bm{\mathcal{Z}}_{t_{c}}({\bf r}).
\end{equation}
The field $\bm{\mathcal{E}}_{ee}({\bf r})$ is obtained via
Coulomb's law from its  quantal source $g({\bf r}{\bf r}^{\prime})$, the pair-correlation density. Thus,
\begin{equation}
\bm{\mathcal{E}}_{ee}({\bf r})=\int
\frac{g({\bf r}{\bf r}^{\prime})
({\bf r}-{\bf r}^{\prime})}{|{\bf r}-{\bf r}^{\prime}|^{3}}
d{\bf r}^{\prime},
\end{equation}
where $g({\bf r}{\bf r}^{\prime}) = <\Psi|\hat{P}({\bf r}{\bf
r}^{\prime})|\Psi>/\rho({\bf r})$, and $\hat{P}({\bf r}{\bf
r}^{\prime})=\sum_{i,j}^{\prime} \delta({\bf r}_{i}-{\bf
r})\delta({\bf r}_{j}-{\bf r}^{\prime})$
 the Hermitian pair-correlation operator. The
pair-correlation density may be further  separated into
its local(static) and nonlocal (dynamic) components as
\begin{eqnarray}
g({\bf r}{\bf r}^{\prime})&=& \rho({\bf r}^{\prime})+
\rho_{xc}({\bf r}{\bf r}^{\prime})\\
&=&\rho({\bf r}^{\prime})+\rho_{x}({\bf r}{\bf r}^{\prime})+
\rho_{c}({\bf r}{\bf r}^{\prime}),
\end{eqnarray}
where the sources $ \rho_{xc}({\bf r}{\bf r}^{\prime})$,
$\rho_{x}({\bf r}{\bf r}^{\prime})$, and
$\rho_{c}({\bf r}{\bf r}^{\prime})$ are the Fermi-Coulomb,
Fermi, and Coulomb hole charge distributions. The Fermi hole
is defined as $\rho_{x}({\bf r}{\bf r}^{\prime})=
-|\gamma_{s}({\bf r}{\bf r}^{\prime})|^2/2\rho({\bf r})$,
 and the Coulomb hole is defined via Eqs.(11) and (12).
 The sum rules satisfied by these charge distributions
 are $\int \rho_{xc}({\bf r}{\bf r}^{\prime})
 d{\bf r}^{\prime}=-1$; $\int \rho_{x}({\bf r}
 {\bf r}^{\prime}) d{\bf r}^{\prime}=-1$;
 $\rho_{x}({\bf r}{\bf r}^{\prime})\leq 0$; $\rho_{x}({\bf r}{\bf r})=
 -\rho({\bf r})/2$, and  $\int \rho_{c}({\bf r}{\bf r}^{\prime})
 d{\bf r}^{\prime}=0$. With the above separation,
 the electron-interaction field may
 then be written in terms of its components as
\begin{eqnarray}
\bm{\mathcal{E}}_{ee}({\bf r})&=&   \bm{\mathcal{E}}_{H}({\bf r})
 + \bm{\mathcal{E}}_{xc}({\bf r})\\
&=& \bm{\mathcal{E}}_{H}({\bf r})+\bm{\mathcal{E}}_{x}({\bf r})
+ \bm{\mathcal{E}}_{c}({\bf r}),
\end{eqnarray}
where the Hartree $\bm{\mathcal{E}}_{H}({\bf r})$, Pauli-Coulomb
 $\bm{\mathcal{E}}_{xc}({\bf r})$, Pauli
 $\bm{\mathcal{E}}_{x}({\bf r})$, and Coulomb
  $\bm{\mathcal{E}}_{c}({\bf r})$ fields are due to
  their respective quantal sources $\rho({\bf r})$,
  $ \rho_{xc}({\bf r}{\bf r}^{\prime})$, $\rho_{x}({\bf r}{\bf r}^{\prime})$,
  and $\rho_{c}({\bf r}{\bf r}^{\prime})$.\\

The Correlation-Kinetic field $\bm{\mathcal{Z}}_{t_{c}}({\bf r})$
is the difference of the kinetic fields $\bm{\mathcal{Z}}({\bf r})$
and $\bm{\mathcal{Z}}_{s}({\bf r})$ of the interacting and noninteracting systems, respectively:
\begin{equation}
\bm{\mathcal{Z}}_{t_{c}}({\bf r})=
\bm{\mathcal{Z}}_{s}({\bf r})-\bm{\mathcal{Z}}({\bf r}),
\end{equation}
where $\bm{\mathcal{Z}}({\bf r})={\bf z}({\bf r};
[\gamma])/\rho({\bf r})$
and $\bm{\mathcal{Z}}_{s}({\bf r})={\bf z}_{s}({\bf r};
[\gamma_{s}])/\rho({\bf r})$. The quantal sources
of the fields $\bm{\mathcal{Z}}({\bf r})$ and
$\bm{\mathcal{Z}}_{s}({\bf r})$
are the single particle and Dirac density matrices.
The kinetic `force' ${\bf z}({\bf r};[\gamma])$ is defined in terms
of its components as $z_{\alpha}({\bf r};
[\gamma])=2\sum_{\beta}\partial t_{\alpha\beta}({\bf r};
[\gamma])/\partial r_{\beta}$, where  $t_{\alpha\beta}({\bf r};
[\gamma])=\frac{1}{4}[\partial^{2}/\partial r_{\alpha}^{\prime}
\partial r_{\beta}^{\prime\prime}+\partial^{2}/\partial r_{\beta}^
{\prime}
\partial r_{\alpha}^{\prime\prime}]\gamma
({\bf r}^{\prime}{\bf r}^{\prime\prime})|_{{\bf r}^{\prime}=
{\bf r}^{\prime\prime}
={\bf r}}$ is the kinetic energy tensor. The field ${\bf z}_{s}({\bf r};
[\gamma_{s}])$ is similarly defined in terms of the S-system tensor
$t_{\alpha\beta,s}({\bf r};[\gamma_{s}])$.\\

The Hartree field $\bm{\mathcal{E}}_{H}({\bf r})$ is conservative,
and $\nabla \times \bm{\mathcal{E}}_{H}({\bf r})=0$. This is because
 its source $\rho({\bf r})$ is a static charge, and the field may
 consequently be written as $\bm{\mathcal{E}}_{H}({\bf r})=-\nabla W_{H}({\bf r})$,
 where $W_{H}({\bf r})=\int d{\bf r'} \rho({\bf r'})/|{\bf r}-{\bf r'}|$. The
 fields $\bm{\mathcal{E}}_{xc}({\bf r})$, $\bm{\mathcal{E}}_{x}({\bf r})$, and
  $\bm{\mathcal{E}}_{c}({\bf r})$
 are in general not conservative as their sources
  are nonlocal.
  The sum of the fields
   $\bm{\mathcal{E}}_{xc}({\bf r})+\bm{\mathcal{Z}}_{t_{c}}({\bf
   r})$    and $
   \bm{\mathcal{E}}_{x}({\bf r})+\bm{\mathcal{E}}_{c}({\bf r})+\bm{\mathcal{Z}}_{t_{c}}({\bf r})$  are always conservative.\\

For systems of symmetry such that the component fields
$\bm{\mathcal{E}}_{ee}({\bf r})$ and
$\bm{\mathcal{Z}}_{t_{c}}({\bf r})$ are separately conservative,
the potential energy $v_{ee}({\bf r})$ may be expressed as
the sum of the separate work done in these fields. Thus
\begin{eqnarray}
    v_{ee}({\bf r})&=& W_{ee}({\bf r})+W_{t_c}({\bf r})\\
    &=&W_{H}({\bf r})+W_{xc}({\bf r})+W_{t_{c}}({\bf r})\\
    &=& W_{H}({\bf r})+W_{x}({\bf r})+W_{c}({\bf r})+W_{t_{c}}({\bf r}),
\end{eqnarray}
where $W_{ee}({\bf r})$, $W_{H}({\bf r})$, $W_{xc}({\bf r})$,
$W_{x}({\bf r})$, $W_{c}({\bf r})$, and $W_{t_{c}}({\bf r})$
are respectively the work done in the fields
$\bm{\mathcal{E}}_{ee}({\bf r})$, $\bm{\mathcal{E}}_{H}({\bf r})$,
$\bm{\mathcal{E}}_{xc}({\bf r})$, $\bm{\mathcal{E}}_{x}({\bf r})$,
$\bm{\mathcal{E}}_{c}({\bf r})$, and
$\bm{\mathcal{Z}}_{t_{c}}({\bf r})$.\\

The ground state energy is
\begin{equation}
E=T_{s}+\int \rho({\bf r})v({\bf r})d{\bf r}+E_{ee}+T_{c},
\end{equation}
where $T_{s} = \left\langle \Phi\{\phi_{i}\}|\hat{T}|\Phi\{\phi_{i}\}
\right\rangle$ is the kinetic energy of the noninteracting Fermions.
The electron-interaction $E_{ee}$ and Correlation-Kinetic $T_{c}$
energies are expressed in terms of the fields
$\bm{\mathcal{E}}_{ee}({\bf r})$
and $\bm{\mathcal{Z}}_{t_{c}}({\bf r})$, respectively,
in integral virial form as
\begin{equation}
E_{ee}=\int d{\bf r}\rho({\bf r})
{\bf r}\cdot\bm{\mathcal{E}}_{ee}({\bf r})
\;\;\;\; {\textstyle and}\;\; T_{c} =\frac{1}{2}\int d{\bf r}
\rho({\bf r}){\bf r}\cdot \bm{\mathcal{Z}}_{t_{c}}({\bf r}).
\end{equation}
These expressions for the energy are valid whether the fields
$\bm{\mathcal{E}}_{ee}({\bf r})$ and
 $\bm{\mathcal{Z}}_{t_{c}}({\bf r})$ are separately conservative or not. Employing Eq.(13)
and (14) in Eq.(20), the energy $E_{ee}$ may be written as a sum of the Hartree
$E_{H}$ and Pauli-Coulomb $E_{xc}$ (or  Pauli $E_{x}$ plus Coulomb $E_{c}$) energies with each component term expressed in integral virial form.\\

Finally, the highest occupied eigenvalue of the S system differential equation
is the negative of the ionization potential: $\epsilon_{m}= -I$.\\

\section{ APPLICATION TO THE HYDROGEN MOLECULE}

\textbf{A. Wave functions, Orbitals, and Density}\\

The purely electronic part of the Hamiltonian for $H_{2}$ in atomic units ($e=m=\hbar=1$) is
\begin{equation}
  \hat{H}= -\frac{1}{2} \nabla_{1}^{2}- \frac{1}{2}
  \nabla_{2}^{2}-\frac {1} {r_{1a}}- \frac {1}{r_{2a}}-\frac {1}{r_{1b}}-\frac {1}{r_{2b}}+\frac {1}
  {r_{12}}
  \end{equation}
  where $1$ and $2$ are the electrons, and $a$ and $b$ are the
  nuclei.
  As the wave function of the molecule in its ground state is unknown, we employ
  the essentially exact $51$-parameter correlated wave function of Kolos-Roothaan  \cite{11}
in our calculations. The symmetric spatial part of the wave
function is
  \begin{equation}
 \Psi({\bf r}_{1} {\bf r}_{2})= exp[-\delta (\xi_{1} +\xi_{2})]
 \sum_{m n j k l} C_{mnjkp} \;[\xi_{1}^{m} \xi_{2}^{n} \eta_{1}^{j}
 \eta_{2}^{k} + \xi_{1}^{n} \xi_{2}^{m} \eta_{1}^{k} \eta_{2}^{j}]
 r_{12}^{p}
 \end{equation}
 with
 \begin{equation}
 \xi_{1}=(r_{1a}+r_{1b})/R; \;\;   \xi_{2}=(r_{2a}+r_{2b})/R ;
 \end{equation}
 \begin{equation}
\eta_{1}=(r_{1a}-r_{1b})/R; \;\;  \eta_{2}=(r_{2a}-r_{2b})/R,
 \end{equation}
 where the variational parameters are $\delta$ and the coefficients
 $C_{mnjkp}$, $r_{12}=|{\bf r}_{1}-{\bf r}_{2}|$, and  $R=2a$ is
the internucleus separation.  The values of the variational
parameters are given in the Appendix. The total energy
 (inclusive of the internuclear potential energy
$V_{nn}=1/R$) is $E_{tot}(H_{2})=-1.174448$ (a.u.) at $a=0.7005$
(a.u.). The kinetic energy $T=-E_{tot}$, and the total potential
energy $E_{ext}+E_{ee}+V_{nn}=-2.348851 $(a.u.). The virial
theorem ratio, which is the ratio of the total potential energy to
twice the total energy, is $0.999981$. The electron interaction
energy component $E_{ee}=0.58737$(a.u.), and the external energy
$E_{ext}=-3.65005$(a.u.). The total energy \cite{18} of the
Hydrogen molecular ion $H_{2}^{+}$ at the equilibrium internuclear
separation of the Hydrogen molecule is
$E_{tot}(H_{2}^{+})|_{a=0.7005}=-0.56998$ (a.u.). Thus, the
ionization potential
of the $H_{2}$ molecule is $I=E_{tot}(H_{2}^{+})|_{a=0.7005}-E_{tot}(H_{2})=0.60447$ (a.u.).\\

 For two electron systems such as the Hooke's atom\cite{19}, Helium atom, or the Hydrogen molecule,
 the orbitals of the S system in its ground (singlet) state that lead to the interacting system
 density are known. These orbitals are $\phi_{i}({\bf r})=\sqrt{\rho({\bf r})/2}, i=1,2$, and are therefore
  known to the same accuracy as the wave function or density.\\

 The density $\rho(0,z)$ along the nuclear bond z-axis is plotted in Fig.1. The density is extremely accurate
 throughout space except at and very near each nucleus. Thus, although on the scale of this figure, it appears that
 the density satisfies the electron-nucleus cusp
 condition\cite{20} exactly,  in fact it does not.\\

 \textbf{B. Fermi-Coulomb, Fermi, and Coulomb Holes }\\

For the $H_{2}$ molecule in its singlet ground state, there are no correlations due to the Pauli
exclusion principle as the two electrons have opposite spin.  However, within the S system framework,
it is customary in local effective potential energy theories to define a Fermi
hole as $\rho_{x}({\bf r} {\bf r'}) = -\rho({\bf r'})/2$.
(This is because the pair-correlation density as determined from the corresponding S system wave
function is $g({\bf r} {\bf r'}) = \rho({\bf r'}) /2$.) \\

In Fig.2 we plot cross-sections through the Fermi-Coulomb
$\rho_{xc}({\bf r} {\bf r'})$, Fermi $\rho_{x}({\bf r} {\bf r'})$,
and Coulomb $\rho_{c}({\bf r} {\bf r'})$ hole sources as a
function of  ${\bf r'} = (0,z')$  for an electron at the origin
${\bf r} = (0,0)$ at the center of the nuclear bond.  (Because of
the cylindrical symmetry of the molecule, cylindrical coordinates
are employed throughout.) The electron position is indicated by
the arrow.  The three charge distributions, of course, have
cylindrical symmetry about the bond axis.  More significantly,
they are symmetrical about the
 electron along the $z'$ axis.  Observe that at the electron position, both the Fermi-Coulomb and
 Coulomb holes exhibit a cusp corresponding to
 the electron-electron cusp condition \cite{20}. (Based on the work of Ref. \cite{21} it is
 known that the wave function does not satisfy this cusp condition exactly. It obviously satisfies it
  to a good degree as evidenced by the figure.)  As expected, at the electron position, the Fermi-Coulomb hole is more negative than the Fermi hole.  Thus, in the region about the electron, the Coulomb hole is negative.  (This is also the case for all the other electron positions considered.)  As both the Fermi-Coulomb and Fermi holes satisfy the same charge conservation sum rule, there must then be regions where the former lies above the latter.  This is clearly evident in the figure.  Hence, in the outer and classically forbidden regions of the molecule, the Coulomb hole is positive.  (The positive part of the Coulomb hole is more clearly evident in the figures that follow.)  The Coulomb hole is both positive and negative as its total charge is zero.  The positive part of the Coulomb hole is an indication that the other electron is equally likely to be in the classically forbidden region on either side of each nucleus. \\

As the Fermi hole is independent of electron position, we now focus on the Fermi-Coulomb and
 Coulomb holes.  In Figs.3-5, we plot the cross-sections of these holes for electron positions
  at ${\bf r} = (0, a/3), {\bf r} = (0, 2a/3), {\bf r} = (0, a)$.  Again, observe the cusp at
   the electron position for both the Fermi-Coulomb and Coulomb holes of each figure.
    Note also how the positive part of the Coulomb hole becomes more pronounced relative to
    the negative part as the electron is moved away from the center of the nuclear bond towards
    one nucleus.  The positive part of the Coulomb hole is also largest about the other nucleus,
    thereby indicating that the second electron is about this nucleus. \\

In Figs.6-8, we plot the Fermi-Coulomb and Coulomb hole
cross-sections for an electron in the classically forbidden region
at ${\bf r} = (0, 2a), {\bf r} = (0, 4a)$, and ${\bf r} = (0,
6a)$. The positive part of the Coulomb hole continues to increase
about the left nucleus at the expense of the negative part as the
electron is moved further from the molecule.  Thus, even for the
asymptotic
 position of an electron at ${\bf r} = (0, 6a)$, the other electron is still mainly about the left nucleus.
For all electron positions, the Fermi-Coulomb hole $\rho_{xc}({\bf r} {\bf r'})$ is negative.\\

(We note that the same cross-sections of the Fermi-Coulomb, Fermi, and Coulomb holes for an
 electron position $0.3$ (a.u.) to the left of the right nucleus, which corresponds
 approximately to our Fig.4, has been plotted by Baerends et al \cite{22} in their study
 of the dissociation of the
 molecule.  However, in their figure, the electron-electron cusp in the
  Fermi-Coulomb and Coulomb holes is not present because the wave function employed by these
  authors is a configuration-interaction type wave function.)\\


\textbf{C.  Fields, Potential Energies, and Energies.} \\

The electron-interaction field ${\mathcal E}_{ee}({\bf r})$, and
its Hartree ${\mathcal E}_{H}({\bf r})$ and Pauli-Coulomb
${\mathcal E}_{xc}({\bf r})$ components along the nuclear bond
axis are plotted in Fig.9.
 Observe that these fields all vanish at the center of the bond axis or origin. This is because their
  corresponding sources -- the pair-correlation density $g({\bf r} {\bf r'})$,
  the density $\rho({\bf r})$, and the Fermi-Coulomb hole
  charge $\rho_{xc}({\bf r} {\bf r'} )$ --- are symmetrical
  about the center of the nuclear bond for this electron
  position (see Figs.1 and 2).  The existence (non-zero value) of these
  fields for all other electron positions is a consequence of the fact
   that their sources are \textit{not symmetrical} about the
   electron (see Figs.1 and 3-8).  The fields are
   also all \textit{antisymmetric} about the
   center of the nuclear bond.
    (This is a reflection of the symmetry about
    the x-y plane at the center of the nuclear bond.
    As such the potential energies obtained from these
    fields will be \textit{symmetric} about this point.)
     In the positive half-space, there is a maximum in the
     electron-interaction and Hartree fields, and a minimum in
      the Pauli-Coulomb field.  The Hartree and Pauli-Coulomb fields
      are of the same order of magnitude and opposite in sign.
      This is because their sources, $\rho({\bf r})$ and $\rho_{xc}({\bf r} {\bf r'})$
      respectively, are of the same order of magnitude and opposite in sign.
       Asymptotically, in the z direction these fields decay
       as ${\mathcal E}_{ee}({\bf r}) \sim 1/z^{2} $,
       ${\mathcal E}_{H}({\bf r}) \sim 2/z^{2} $, and
        ${\mathcal E}_{xc}({\bf r}) \sim -1/z^{2}$   as they must \cite{3,9}.  (It is interesting to note that with a slight translation to the right, the plots of the fields in the positive half-space, are strikingly similar to those of the Helium atom \cite{2,9}.)\\

The Pauli ${\mathcal E}_{x}({\bf r})$ and Coulomb ${\mathcal
E}_{c}({\bf r})$  field components of the Pauli-Coulomb field
${\mathcal E}_{xc}({\bf r})$ along the nuclear bond axis are
plotted in Fig.10.  Again, these fields vanish at the origin and
are \textit{antisymmetric} about it. Hence, the corresponding
potential energies obtained from these fields will be symmetric.
In the positive half space, the Pauli field ${\mathcal E}_{x}({\bf
r})$ is negative as its source
 is a negative charge.  The Coulomb field ${\mathcal E}_{c}({\bf r})$, on the other hand, is positive in the
 inter-nuclear region and negative throughout the region beyond the right nucleus.  This structure
 is attributable to the fact that the Coulomb hole has both a positive and negative component.
 Asymptotically, the Pauli field decays as ${\mathcal E}_{x}({\bf r}) \sim -1/z^{2} $,
 whereas the Coulomb field  ${\mathcal E}_{c}({\bf r})$ has essentially vanished by about $z = 5$
 (a.u.).
 (Once again in the positive half-space, the structure of these fields when translated slightly to the
  right, is similar to those of the Helium atom.  In particular, we note that the structure of the
   Coulomb holes of the Hydrogen molecule for electron positions $z > a$  (see Figs. 6-8) is very
   similar to those of the Helium atom for electron positions away from its nucleus (see Figs.3,4 of \cite{9}).)
   As is the case for atoms, it turns out that the asymptotic structure along the nuclear
    bond axis of  $\{{\mathcal F}^{eff}({\bf r}) - {\mathcal E}_{H}({\bf r})\} \sim {\mathcal E}_{x}({\bf r})
    \sim -1/z^{2}$ .  Thus, the asymptotic structure of the electron-interaction potential
    energy $v_{ee}({\bf r})$ minus the Hartree potential energy $W_{H}({\bf r})$ is again due
    to Pauli correlations: $\{v_{ee}({\bf r}) - W_{H}({\bf r})\} \sim W_{x}({\bf r}) \sim - 1/z $ as shown
    in Fig. 11. \\

Since in the S system description of two electron systems ${\mathcal E}_{x}({\bf r}) = - {\mathcal E}_{H}({\bf r})/2$,
 the curl of the Fermi field along the nuclear bond z axis
  direction vanishes: $\nabla \times {\mathcal E}_{x}({\bf r})|_{z}= 0$, as it does in all directions.
   Hence, the work done $W_{x}(0,z)$ plotted in Fig. 11 is path independent.
   Along the nuclear bond axis, however, the $\nabla \times {\mathcal E}_{c}({\bf r})|_{z}\neq 0$
   and $\nabla \times {\mathcal Z}_{t_{c}}({\bf r})|_{z}\neq 0$ .  But in this and all directions,
   the curl of the sum of the fields ${\mathcal E}_{c}({\bf r})$ and  ${\mathcal Z}_{t_{c}}({\bf r})$
   vanishes: $\nabla \times [{\mathcal E}_{c}({\bf r})  + {\mathcal Z}_{t_{c}}({\bf r})])|_{z} = 0$.
   Therefore, the work done in the sum of these fields in all directions, and hence along the nuclear
    bond axis $v_{c}(0,z) = W_{c}(0,z) + W_{t_{c}}(0,z)$ is path independent.
    The calculation of the potential energy $v_{c}(0,z)$ is straightforward.  However,
    our use of the Kolos-Roothaan wave function, in spite of its accuracy, leads to $v_{c}(0,z)$ being
     singular at the nucleus.  This occurs due to the component $W_{t_{c}}(0,z)$ that requires a
     cancellation of the kinetic fields of the interacting and noninteracting systems.
     The underlying reason for the singularity, however, is that the wave function does not
     satisfy the \textit{electron-nucleus} cusp condition exactly.  In a recent paper \cite{23},
     we have proved by employing the integral form of the electron-nucleus cusp condition \cite{24}, that
     in local effective potential energy theories and for arbitrary symmetry,
     the potential energy $v_{ee}({\bf r})$ is finite at the nucleus. Furthermore,
     it is shown that this finiteness is a direct consequence of the satisfaction of the
      electron-nucleus cusp condition by the Schrodinger wave
      function. (As a consequence, for example, this potential
      energy is singular at each nucleus when determined either
      from Gaussian geminal \cite{23} or configuration interaction
      \cite{25} wave functions.)  Hence, in order to obtain $v_{c}(0,z)$, we have employed
       our calculated results in regions other than near the
       nucleus, and smoothed the curve through each nucleus.( A comparison of our results with
       the work of Gritsenko \emph{et al}\cite{26} who in their self-consistent calculations assumed
       $v_{ee}({\bf r})$ to be finite at the nucleus show the two curves to be indistinguishable throughout space.)
        The potential
      energy $v_{c}(0,z)$ is plotted in Fig.12.  Observe that $v_{c}(0,z)$, and thus the sum of
      the Coulomb and Correlation-Kinetic potential energies is an order of magnitude
       smaller than $W_{x}(0,z)$, the Pauli contribution.  The potential energy $v_{c}(0,z)$ has
       considerable structure, is symmetric about the origin, and is mainly positive,
       indicating thereby that its principal contribution is Correlation-Kinetic.
        (Recall that the Coulomb field is principally negative in the right-half space
        (see Fig. 10) so that the Coulomb potential energy $W_{c}({\bf r})$ is negative.) The plot of
        $v_{c}(0,z)$ translated to the right nucleus is very similar in shape and magnitude to the
        corresponding potential energy $v_{c}({\bf r})$ of the Helium atom (see Fig. 4 of \cite{2}). \\

To obtain a quantitative sense of the separate Coulomb and Correlation-Kinetic
contributions to $v_{c}(0,z)$, we plot in both Figs.11 and 12 the work done
 $W_{c}(0,z)$ along the path of the nuclear bond in the force of the Coulomb
  field ${\mathcal E}_{c}({\bf r})$.  From $v_{c}(0,z)$ and  $W_{c}(0,z)$ we
  obtain $W_{t_{c}}(0,z)$ which is also plotted in Fig. 12.  The corresponding
   Correlation-Kinetic field $Z_{t_{c}}(0,z)$ is shown in Fig.13. Note that
    this field too is antisymmetric about the origin. Once again, there is a striking similarity
     between the plots of  $W_{c}(0,z)$, $Z_{t_{c}}(0,z)$, and $W_{t_{c}}(0,z)$ when translated to
     the right nucleus to those of the corresponding properties of the Helium atom \cite{2}.
     The Coulomb correlation part  $W_{c}(0,z)$ is negative throughout space and vanishes by about
      $z = 5 $ (a.u.).  The Correlation-Kinetic piece $W_{t_{c}}(0,z)$ is throughout
      positive and asymptotically decays more slowly. The field $Z_{t_{c}}(0,z)$ is principally
      positive throughout space. Thus, the Correlation-Kinetic energy $T_{c}$ is
      positive: $T = 1.1745$ (a.u.), $T_{s} = 1.1414$ (a.u.), $T_{c} = 0.0331$ (a.u.)
      (The corresponding value of $T_{c}$ for the Helium atom is $0.0365$ (a.u.)\cite{9}).
      (We note that the $W_{c}(0,z)$ and $W_{t_{c}}(0,z)$  do not each separately represent a potential energy.
       Their sum which is $v_{c}(0,z)$ does.)\\

\section {   \textbf{CONCLUDING REMARKS}}

This paper is the first application of the Q-DFT quantal source and field perspective
 to a molecule, and much has been learned as explained in the previous section.
 The symmetry of the $H_{2}$ molecule dictates that the individual fields ${\mathcal E}_{x} ({\bf r})$,
  ${\mathcal E}_{H} ({\bf r})$, ${\mathcal E}_{c} ({\bf r})$, ${\mathcal Z}_{t_{c}}(r)$ representative of the
  Pauli and Coulomb correlations, and Correlation-Kinetic effects respectively, must each be antisymmetric
   about the center of the nuclear bond.  The corresponding electron-interaction potential
    energy $v_{ee}({\bf r})$ representative of these correlations as determined by the work done
     in the force of these fields is then symmetric about this point as also dictated by the
     symmetry of the molecule.  The potential energy $v_{ee}({\bf r})$ is also finite at each nucleus,
     as must be the case \cite{23}. The Hartree ${\mathcal E}_{H} ({\bf r})$  and Pauli ${\mathcal E}_{x} ({\bf r})$
     fields are the largest in magnitude and opposite in sign, the former being positive
      and twice as large as the latter. As such the principal contributions to the electron-interaction
      energy $E_{ee}$ and potential energy $v_{ee}({\bf r})$ are due
      to the Hartree and the Pauli correlation terms. The Coulomb ${\mathcal E}_{c} ({\bf r})$ and
       Correlation-Kinetic ${\mathcal Z}_{t_{c}}(r)$ fields tend to cancel each other,
       so that the contribution of their sum to the potential energy $v_{ee}({\bf r})$ is an
        order of magnitude smaller. However, as the potential energy component $v_{c}({\bf r})$ representing
        the sum of these correlations is principally positive (see Fig. 12), it is evident that the
         Correlation-Kinetic effects are more significant. They are also more significant asymptotically,
          where the Coulomb correlation contributions to the potential energy vanish. Thus,
          Correlation-Kinetic effects play an important role in local effective potential energy
           theories of the $H_{2}$ molecule.
           We further note that in the construction of approximate KS-DFT `exchange-correlation' and correlation energy
           functionals $E^{KS}_{xc}[{\bf r}]$ and $E^{KS}_{c}[{\bf r}]$ for molecules, Correlation-Kinetic
            effects must be
           incorporated if an accurate S system representation of molecules is to be obtained.\\

           On the basis of the Q-DFT results determined from the
           $H_{2}$ molecule, it is evident that the qualitative
           features of the quantal sources, fields, and potential
           energies for other diatomic  molecules will be similar.
           However, the fields and hence the potential energies of
           these diatomics will have more structure as a
           consequence of the additional  molecular subshells. We
           expect this added structure to be similar to that
           observed in atoms as the number of shells is increased.
           Finally, we note that the accuracy of approximation
           methods within Q-DFT \cite{27} and KS-DFT can be tested
           by comparison with these essentially exact results.\\

We conclude by reiterating the striking similarity between the Q-DFT properties of
the Hydrogen molecule and the Helium atom for electron positions in the positive half space.
 It is interesting that in spite of the presence of a second nucleus, and therefore of a different
  symmetry, the quantal sources and fields representative of the various electron correlations
  in the Hydrogen molecule are so similar to those of the Helium atom. This
   speaks to the commonality of properties of these distinct quantum systems as exhibited within the framework of Q-DFT. \\

\begin{acknowledgements}

We thank Prof. E. J. Baerends and Dr. Myrta Gr{\"u}ning for
sending us the results of Ref. \cite{26}. This work was supported
in part by the Research Foundation of the City University of New
York.
\end{acknowledgements}
\newpage


\appendix
\textbf{APPENDIX:  Wave function parameters}\\
The values of the parameter $\delta$ and the coefficients
$c_{mnjkp}$ for the wave function of Eq.(21) are listed in the
table I.

\begin{table}
\caption{\label{tab:table1} Variational parameters in the
normalized 51-parameter correlated wave function for the ground
state of $H_{2}$\cite{11}.}
\renewcommand{\arraystretch}{0.4}
\begin{tabular}{ll@{\hspace{2mm }}ll@{\hspace{2mm }}l@{\hspace{1mm }}|r}
\hline \hline
\multicolumn{5}{c}{No. of terms}  & $50 $\\

  \multicolumn{5}{r} {$\delta$ =} & $0.995 $\\ \hline
  $\xi_{1}$ & $\eta_{1}$ & $\xi_{2}$ &  $\eta_{2}$ & $ r_{12}$  & \multicolumn{1}{c} {Coefficients} \\ \hline
0  &0 & 0 & 0 & 0 &  $2.065908$ \\
0  &0 & 0 & 2 & 0 &  $1.282032$ \\
0  &0 & 1 & 0 & 0 &  $0.144619$ \\
0  &1 & 0 & 1 & 0 &  $-0.430253$ \\
0  &0 & 0 & 0 & 1 &  $0.787198$ \\

1  &1 & 0 & 1 & 0 &  $-0.235454$ \\
1  &0 & 0 & 2 & 0 &  $0.148273$ \\
0  &0 & 2 & 0 & 0 &  $0.109859$ \\
0  &0 & 0 & 0 & 2 &  $-0.212159$ \\
1  &0 & 1 & 0 & 0 &  $-0.081387$ \\

0  &2 & 0 & 2 & 0 &  $0.182892$ \\
0  &0 & 0 & 2 & 1 &  $0.198555$ \\
0  &0 & 1 & 0 & 1 &  $0.324658$ \\
1  &1 & 1 & 1 & 0 &  $-0.010794$ \\
0  &0 & 1 & 0 & 2 &  $0.077830$ \\

1  &0 & 2 & 0 & 0 &  $-0.055114$ \\
0  &1 & 0 & 1 & 1 &  $0.130714$ \\
0  &1 & 0 & 1 & 2 &  $-0.050854$ \\
1  &0 & 2 & 0 & 1 &  $0.014963$ \\
0  &0 & 2 & 0 & 1 &  $-0.132980$ \\

1  &1 & 1 & 1 & 2 &  $0.000362$ \\
0 &0 & 2 & 0 & 2 &  $0.006992$ \\
1  &0 & 0 & 2 & 1 &  $-0.050940$ \\
1  &1 & 1 & 1 & 1 &  $0.018027$ \\
1  &0 & 1 & 0 & 1 &  $0.017554$ \\

0  &0 & 0 & 2 & 2 &  $-0.014601$ \\
1  &0 & 1 & 0 & 2 &  $-0.015172$ \\
1  &0 & 0 & 2 & 2 &  $0.012656$ \\
1  &2 & 3 & 0 & 0 &  $-0.000202$ \\
2  &0 & 3 & 0 & 0 &  $-0.000856$ \\

0  &0 & 1 & 2 & 0 &  $-0.009469$ \\
0  &0 & 3 & 0 & 0 &  $0.036963$ \\
1  &0 & 1 & 2 & 0 &  $-0.022325$ \\
0  &1 & 2 & 1 & 0 &  $0.053233$ \\
1  &0 & 3 & 0 & 0 &  $0.004690$ \\

1  &2 & 1 & 2 & 0 &  $0.004707$ \\
1  &1 & 2 & 1 & 0 &  $-0.017531$ \\
0  &2 & 3 & 0 & 0 &  $0.017270$ \\
3  &0 & 3 & 0 & 0 &  $0.000082$ \\
2  &1 & 2 & 1 & 0 &  $0.000031$ \\

0  &0 & 1 & 2 & 1 &  $0.094436$ \\
0  &0 & 3 & 0 & 1 &  $0.001789$ \\
0  &0 & 3 & 0 & 2 &  $-0.000394$ \\
0  &0 & 1 & 2 & 2 &  $-0.004475$ \\
2  &0 & 3 & 0 & 1 &  $-0.000121$ \\

1 &0 & 1 & 2& 1 &  $-0.014893$ \\
2  &0 & 3 & 0 & 2 &  $0.000011$ \\
1 &0 & 1 & 2& 2 &  $0.001016$ \\
0  &2 & 3 & 0 & 1 &  $-0.003443$ \\
0  &2 & 3 & 0 & 2 &  $0.000225$ \\ \hline \hline

\end{tabular}
\end{table}


%
\begin{figure}
\includegraphics[bb= 1 2 500 658, angle=90,scale=0.8]{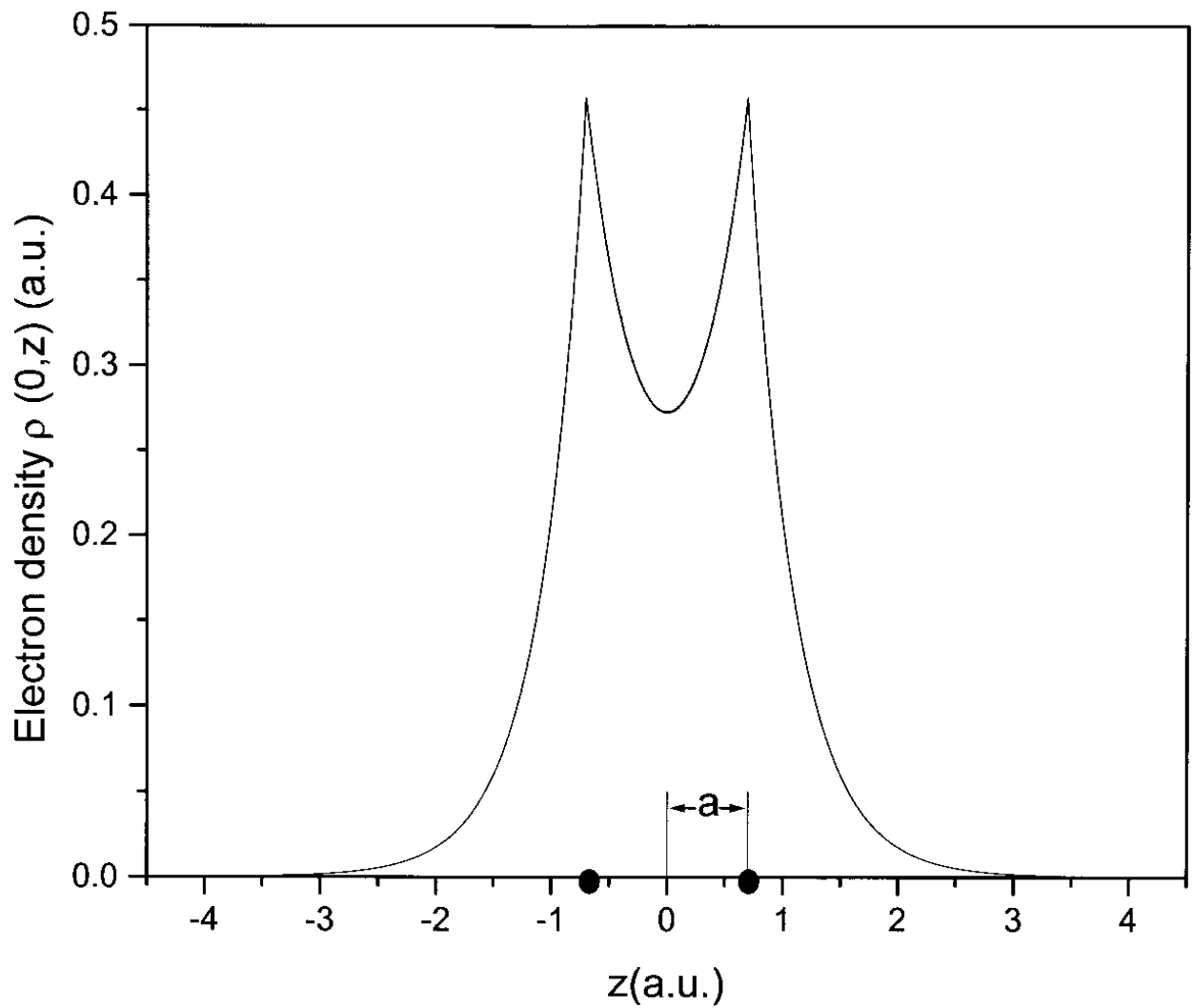}
\caption{ The electron density $\rho(0,z)$ of the hydrogen
molecule along the nuclear bond axis in atomic units (a.u.).  The
nuclei are on the axis at $a = ± 0.7005 $ (a.u.) indicated by the
two dots.\label{}}
\end{figure}

\begin{figure}
\includegraphics[bb=1 1 504 627, angle=90.8,scale=0.8]{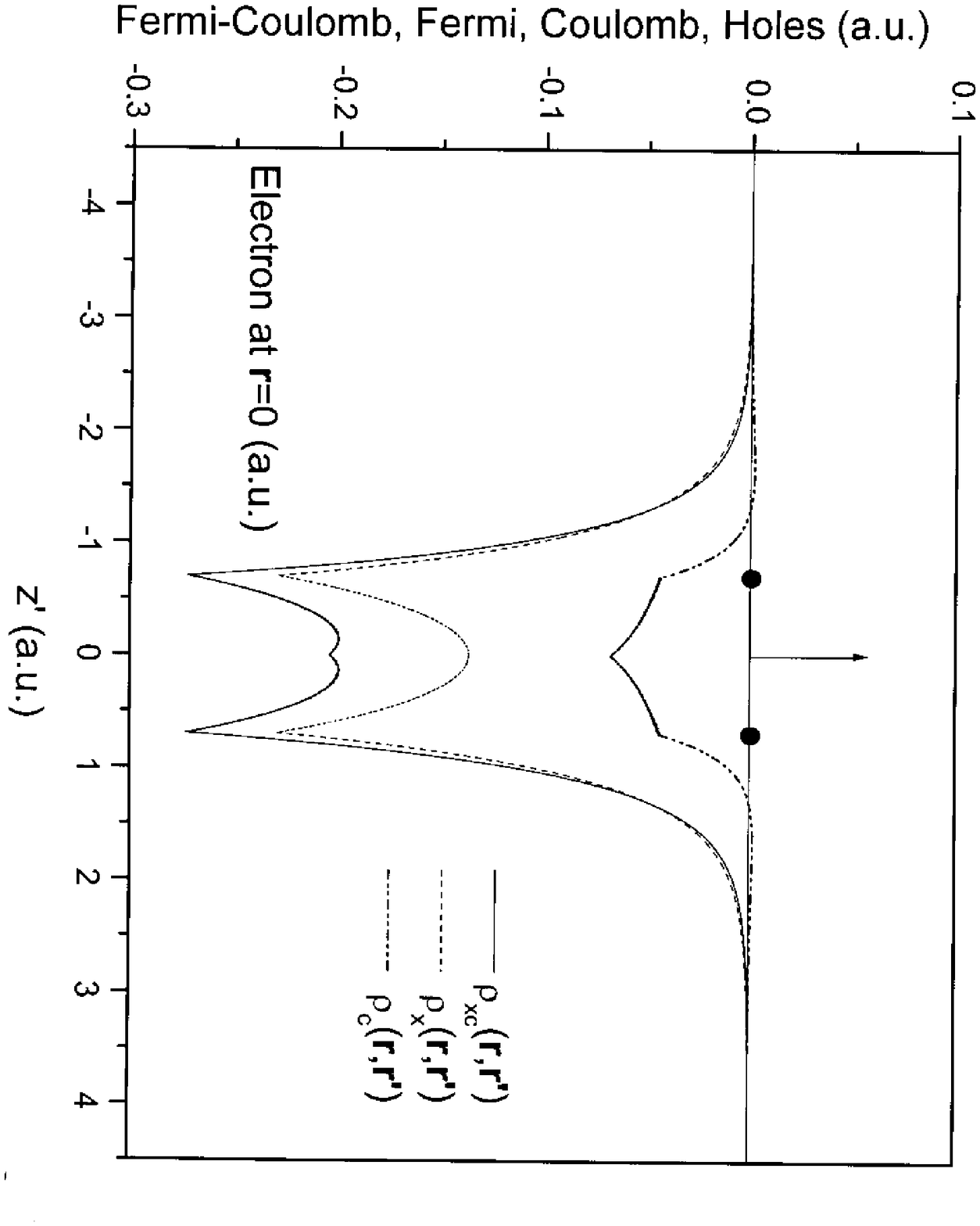}
\caption{ Cross-sections of the Fermi-Coulomb $\rho_{xc}({\bf r} {\bf r'})$, Fermi $\rho_{x}({\bf r} {\bf r'})$, and Coulomb   $\rho_{c}({\bf r} {\bf r'})$        holes along the nuclear bond axis for an electron at the center ${\bf r} = (0,0)$ of the bond.  The electron position is indicated by the arrow.\label{}}
\end{figure}

\begin{figure}
\includegraphics[bb=1 1 504 679,angle=90.5,scale=0.75]{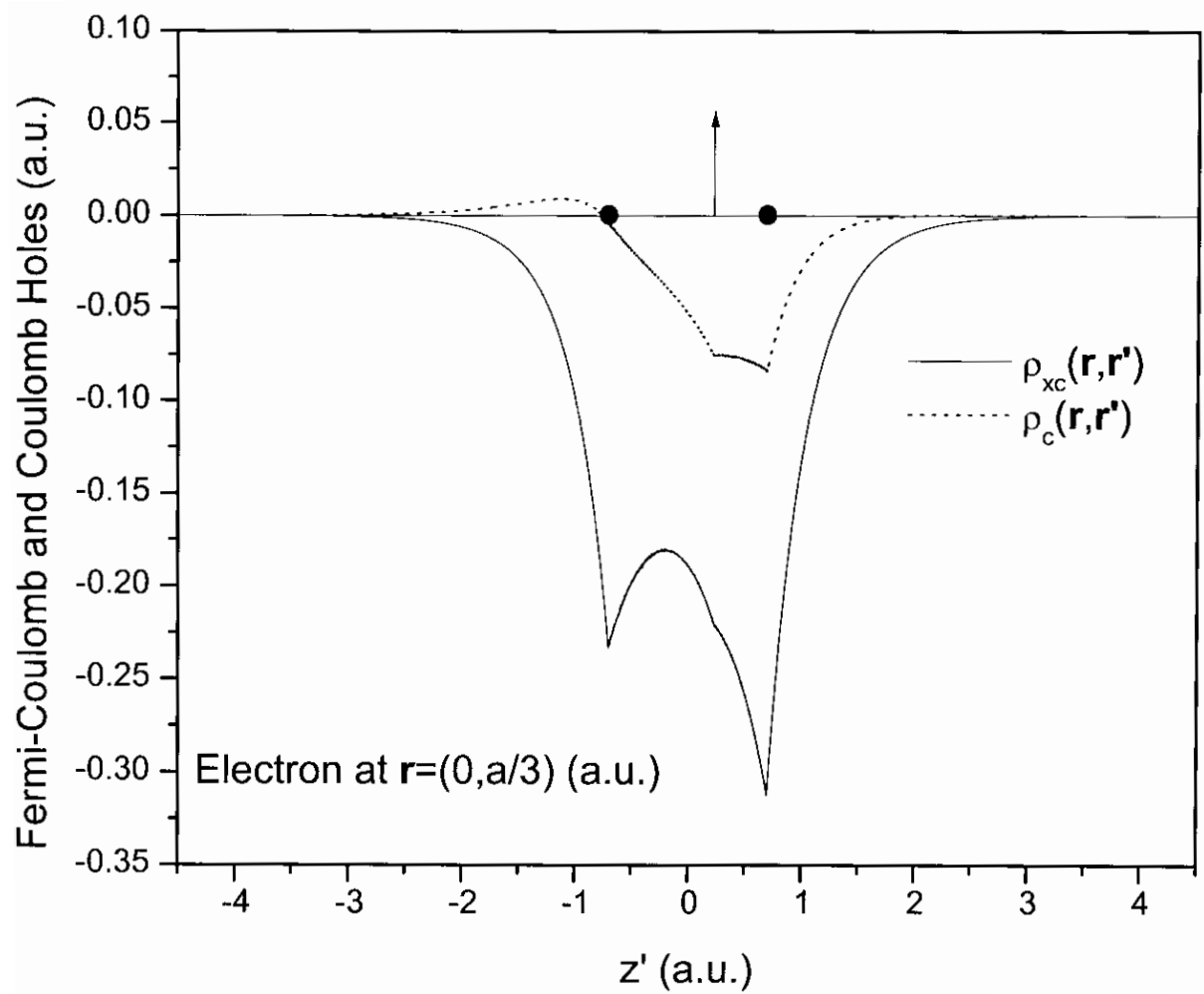}
\caption{ Cross-sections of the Fermi-Coulomb $\rho_{xc}({\bf r} {\bf r'})$  and Coulomb   $\rho_{c}({\bf r} {\bf r'})$   holes along the nuclear bond axis for an electron at the center ${\bf r} = (0,a/3)$ of the bond with  the electron position  indicated by the arrow.\label{}}
\end{figure}

\begin{figure}
\includegraphics[bb=1 1 502 646,angle=90,scale=0.45]{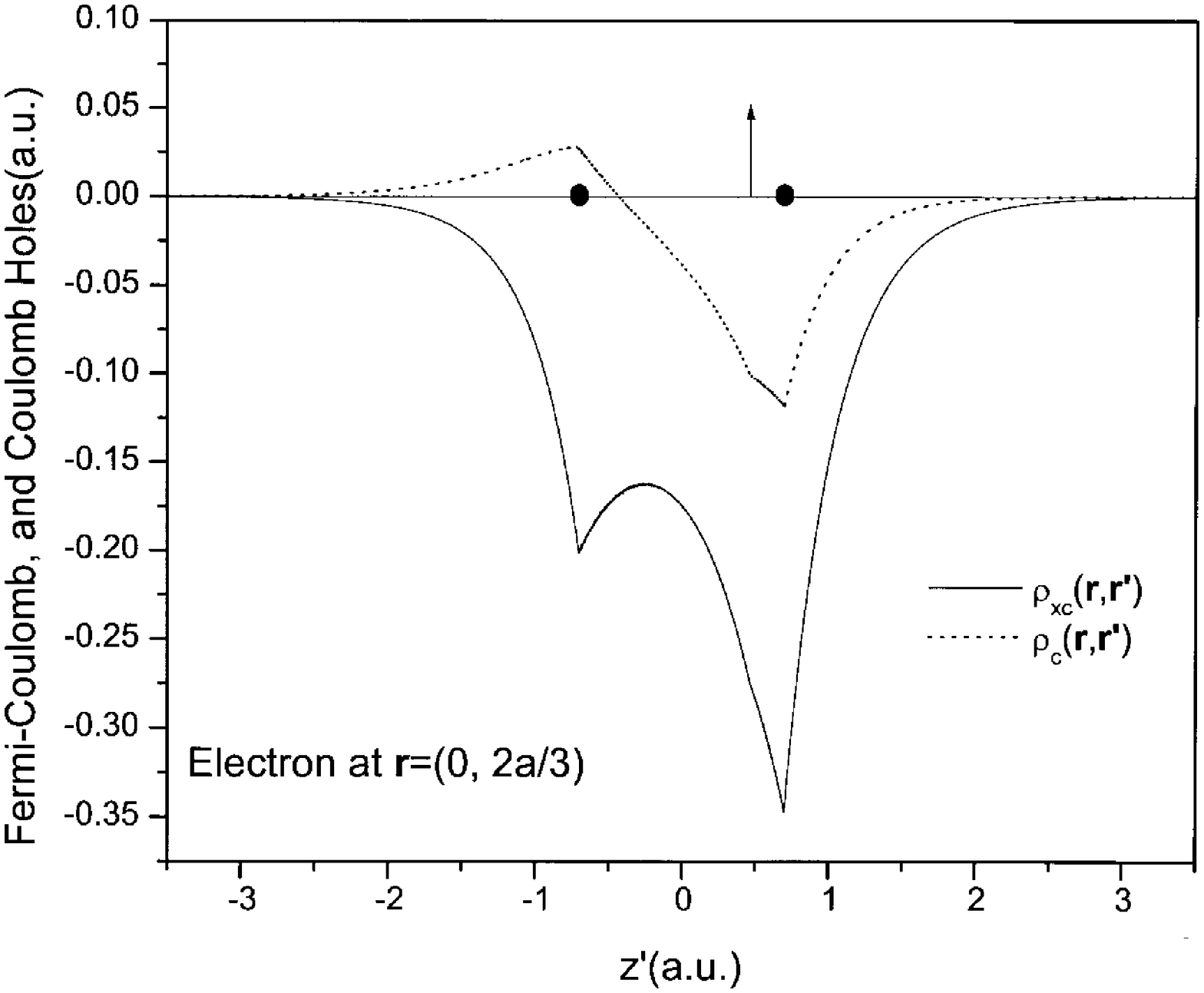}
\caption{ The same as in Fig.3, but with the electron at ${\bf r}
= (0, 2a/3)$.\label{}}
\end{figure}

\begin{figure}
\includegraphics[bb=0 0 50 676,angle=90,scale=0.45]{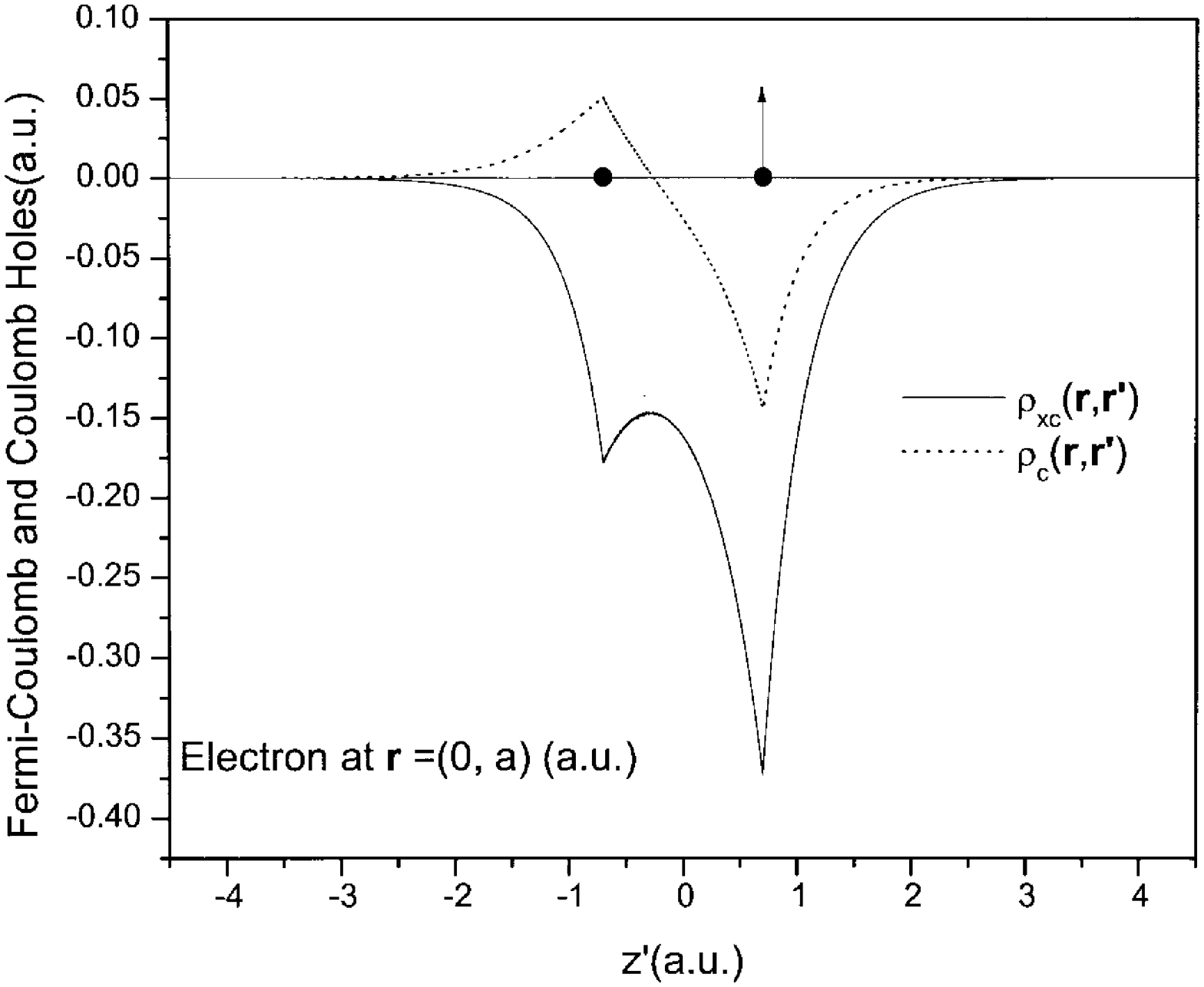}
\caption{ The same as in Fig.3, but with the electron at ${\bf r}
= (0, a)$.\label{}}
\end{figure}

\begin{figure}
\includegraphics[bb=1 1 502 630,angle=90,scale=0.8]{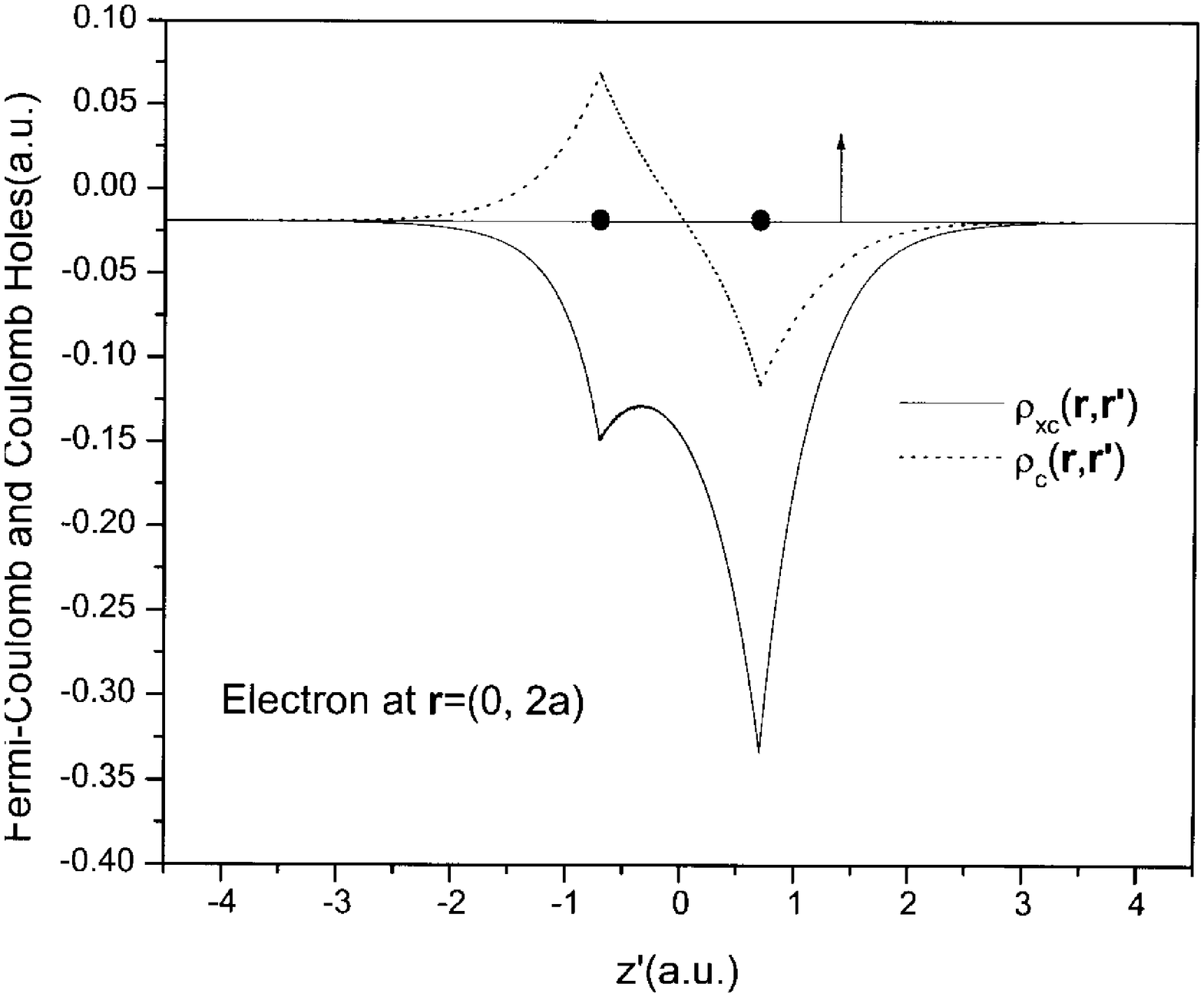}
\caption{ The same as in Fig.3, but with the electron at ${\bf r}
= (0, 2a)$.  \label{}}
\end{figure}

\begin{figure}
\includegraphics[bb=1 1 491 633,angle=90,scale=0.8]{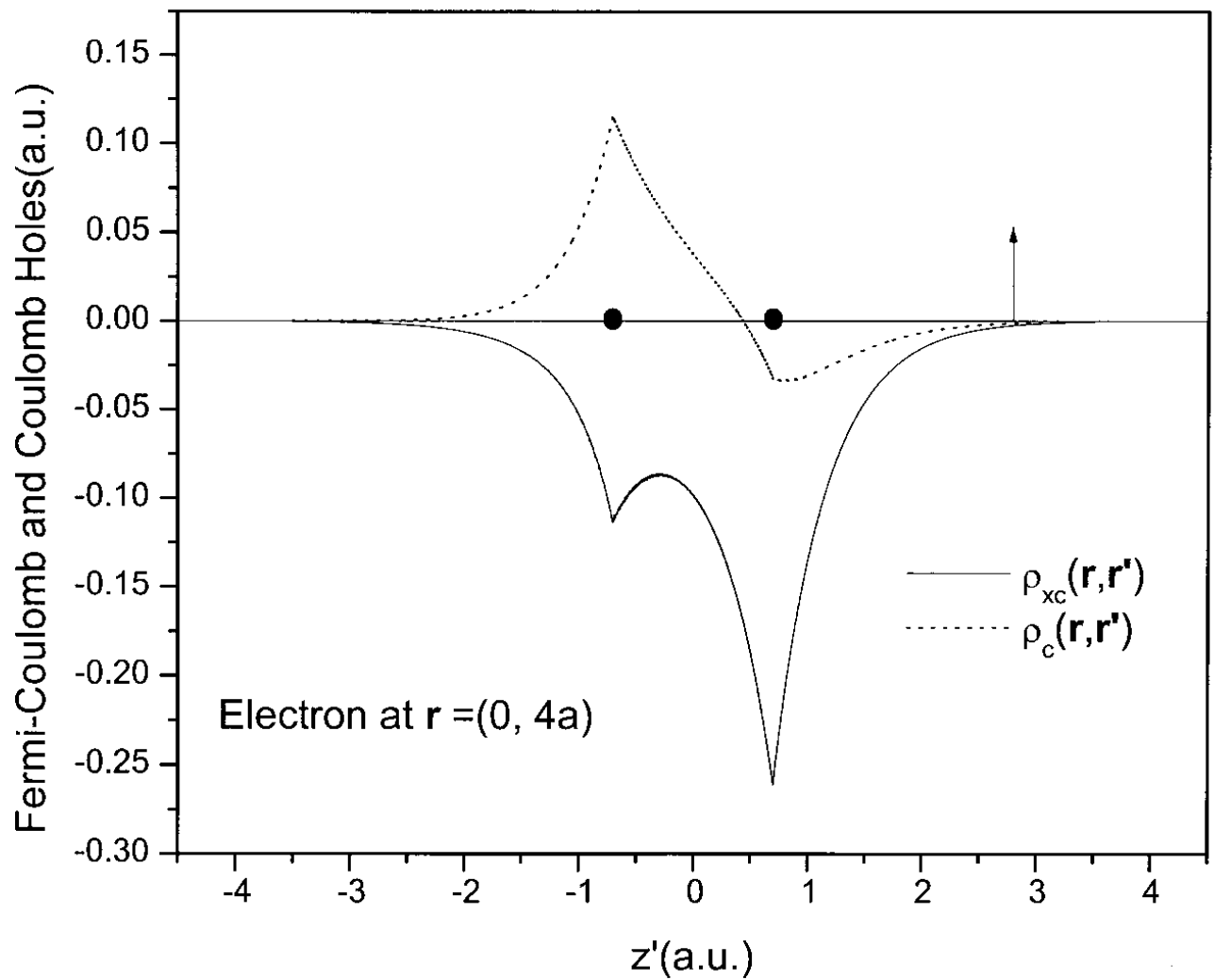}
\caption{ The same as in Fig.3, but with the electron at ${\bf r}
= (0, 4a)$.  \label{}}
\end{figure}

\begin{figure}
\includegraphics[bb=1 1 504 628,angle=90.4,scale=0.8]{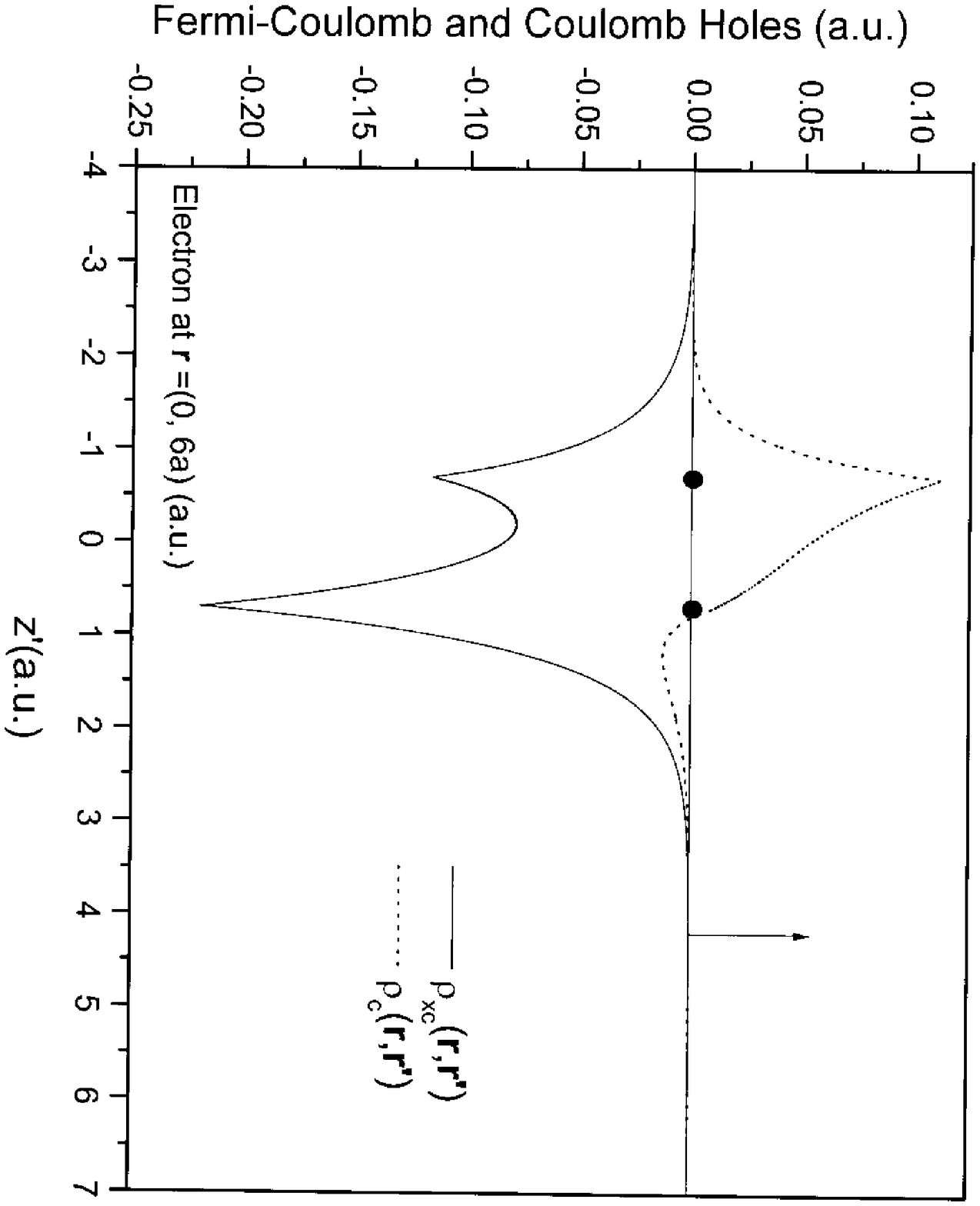}
\caption{ The same as in Fig.3, but with the electron at ${\bf r}
= (0, 6a)$.  \label{}}
\end{figure}

\begin{figure}
\includegraphics[bb=1 1 495 617,angle=90,scale=0.8]{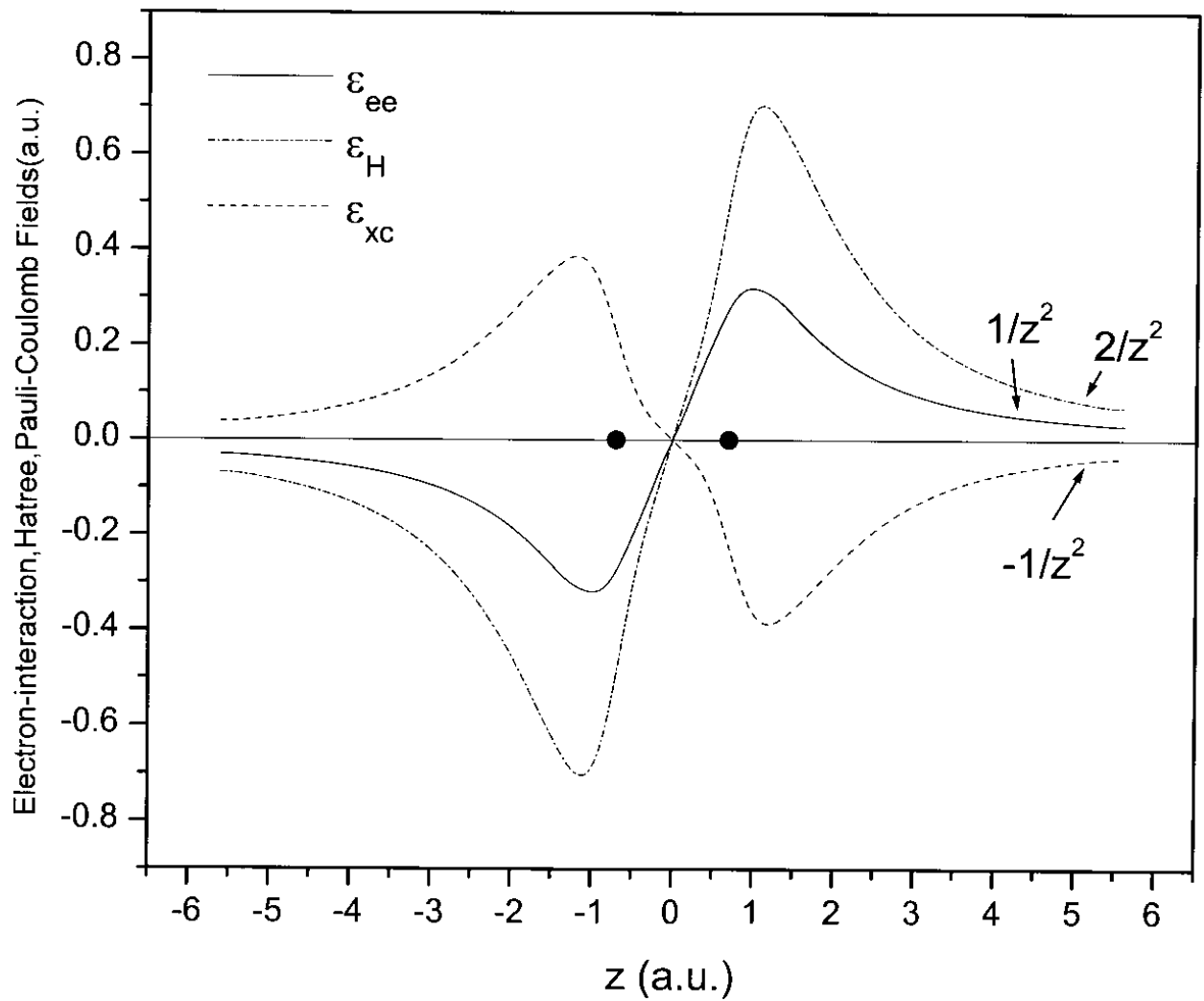}
\caption{ The electron-interaction ${\mathcal E}_{ee}(0,z)$ field, and its Hartree ${\mathcal E}_{H}(0,z)$
 and Pauli-Coulomb ${\mathcal E}_{xc}(0,z)$ components along the nuclear bond axis.  \label{}}
\end{figure}

\begin{figure}
\includegraphics[bb=1 1 495 620,angle=90,scale=0.8]{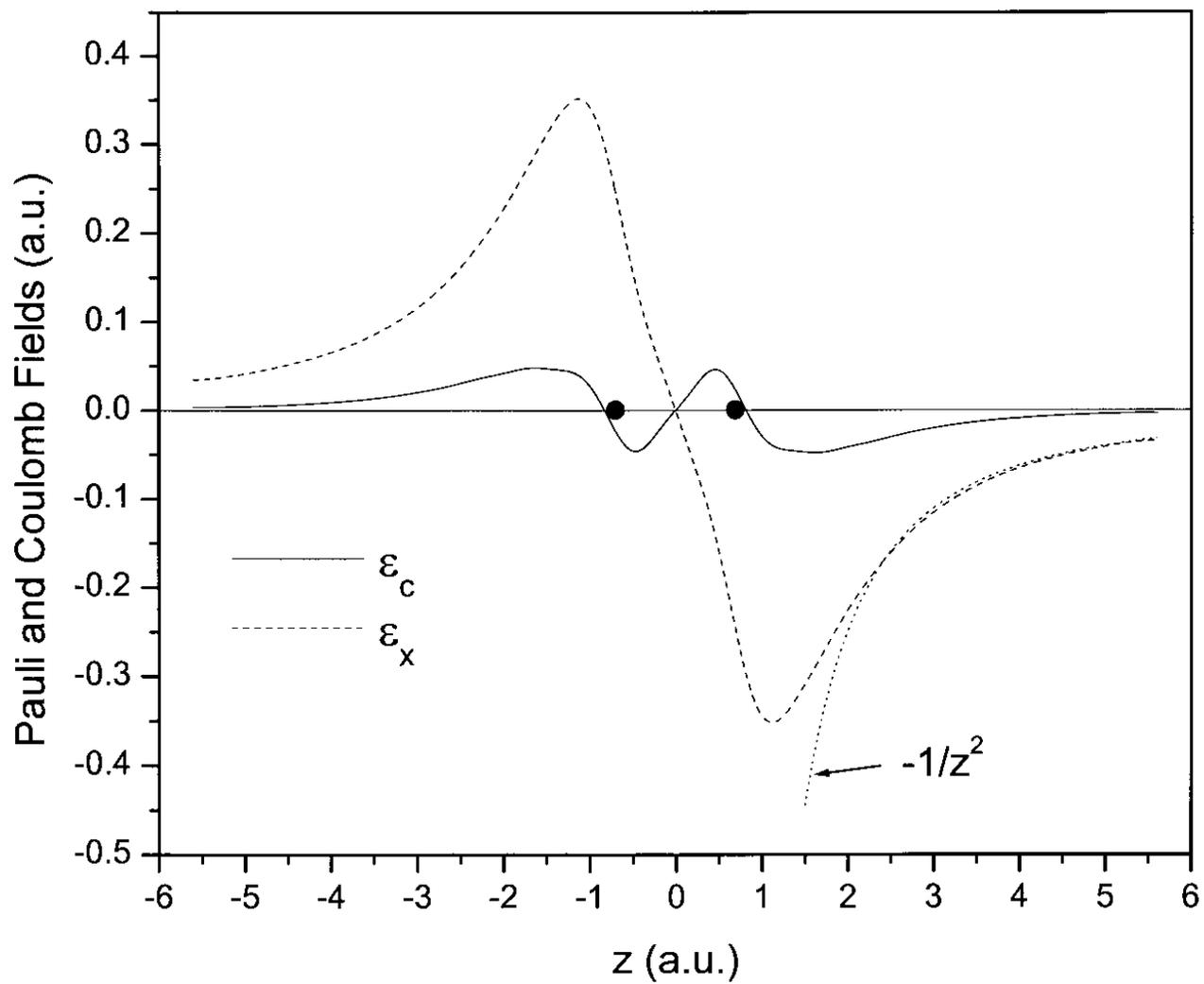}
\caption{ The Pauli ${\mathcal E}_{x}(0,z)$ and Coulomb ${\mathcal E}_{c}(0,z)$ fields
along the nuclear bond axis.  The function $- 1/z^{2}$ is also plotted. \label{}}
\end{figure}

\begin{figure}
\includegraphics[bb=1 1 495 610,angle=90,scale=0.8]{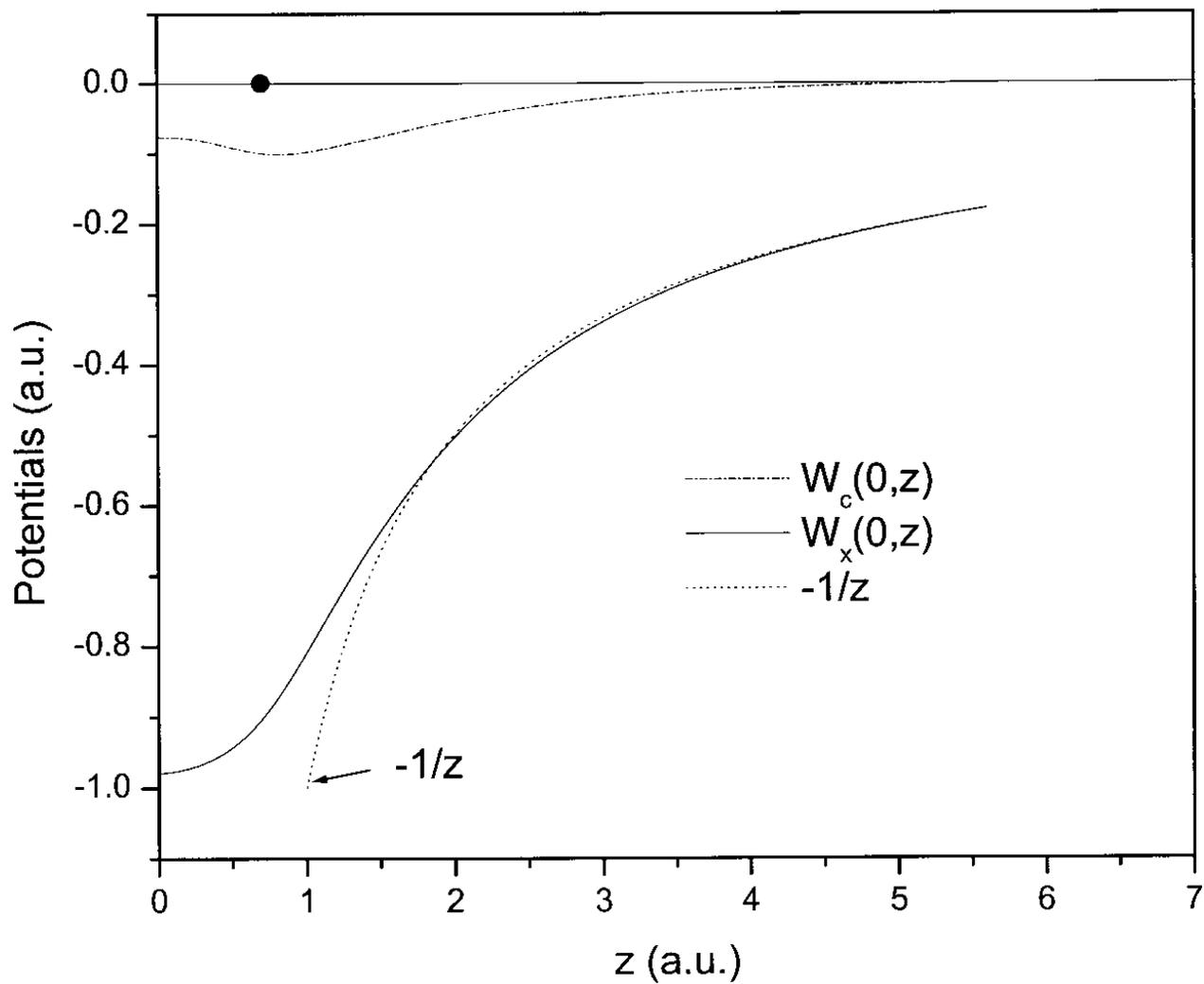}
\caption{ The Pauli potential energy $W_{x}(0,z)$ along the nuclear bond axis.
The work done $W_{c}(0,z)$ in this direction in the force of the Coulomb field ${\mathcal E}_{c}(0,z)$,
and the function $- 1/z$, are also plotted.\label{}}
\end{figure}

\begin{figure}
\includegraphics[bb=1 1 497 608,angle=90,scale=0.8]{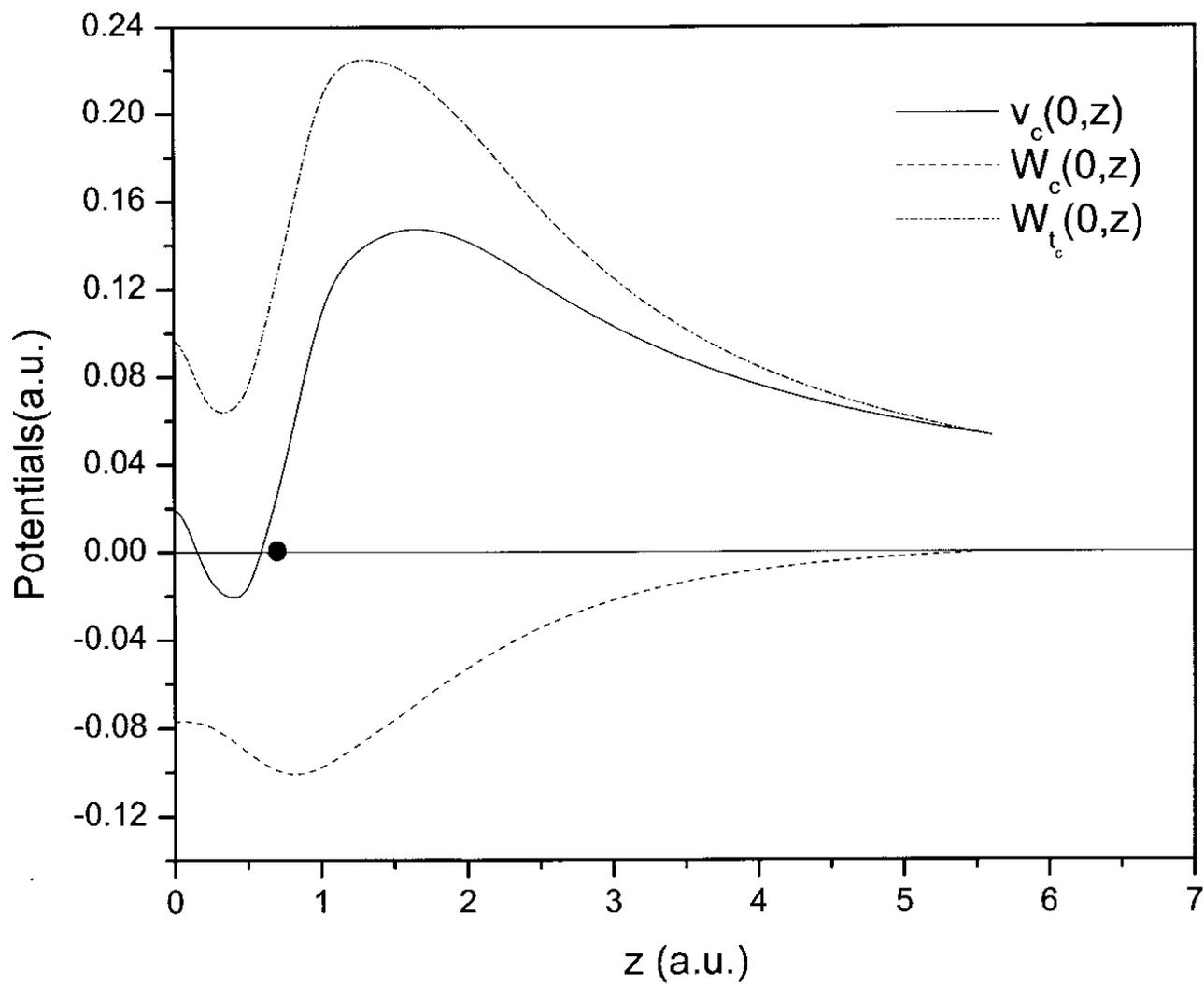}
\caption{ The potential energy $v_{c}(0,z)$, the sum of the
Coulomb and Correlation-Kinetic potential energies, along the
nuclear bond axis.  The work done $W_{c}(0,z)$ of Fig. 11, and the
work done $W_{t_{c}}(0,z)$ in the force of the Correlation-Kinetic
field ${\mathcal Z}_{t_{c}}(0,z)$ of Fig.13, are also
plotted.\label{}}
\end{figure}

\begin{figure}
\includegraphics[bb=1 1 498 611,angle=89.5,scale=0.8]{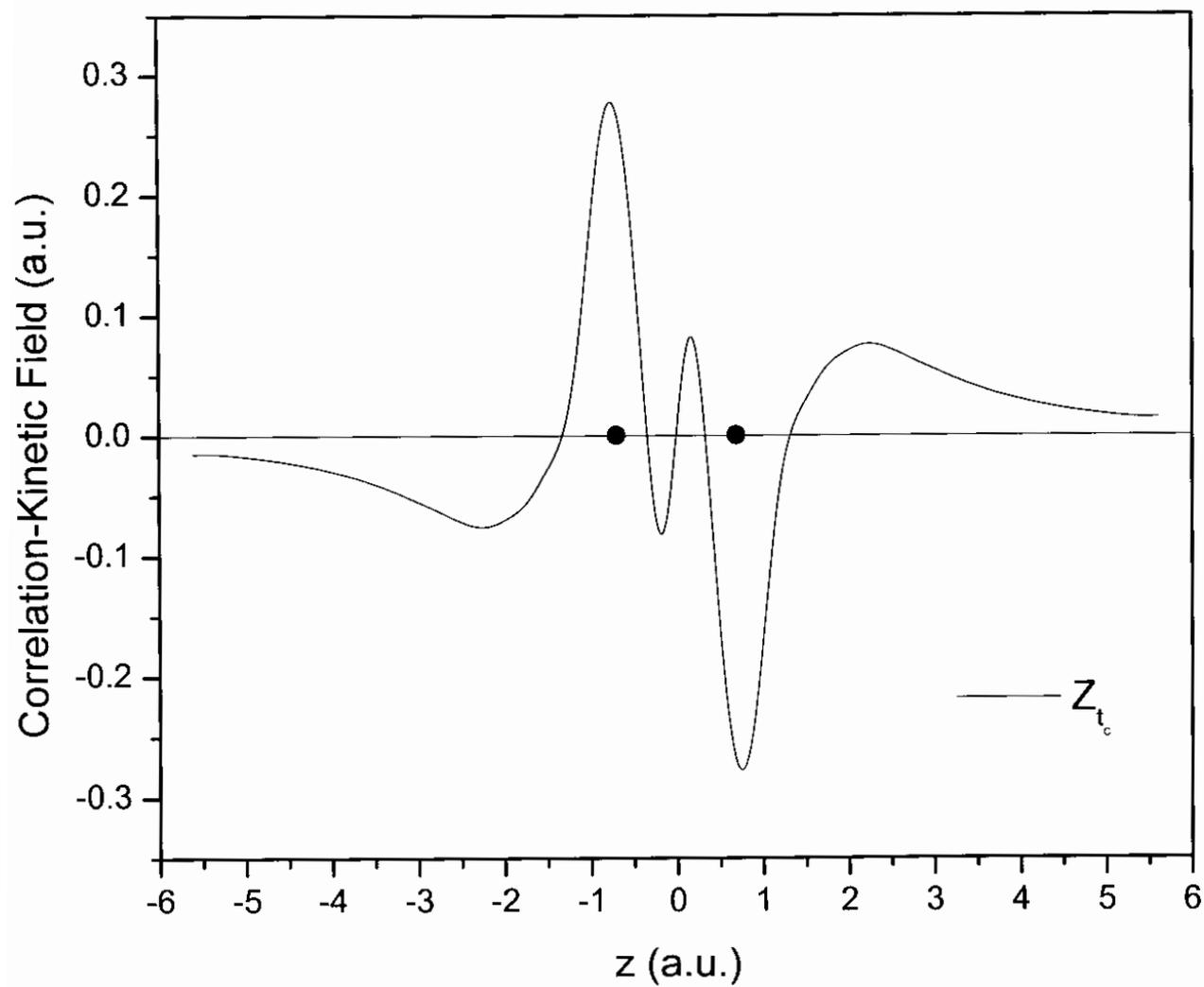}
\caption{ The Correlation-Kinetic field ${\mathcal Z}_{t_{c}}(0,z)$ along the nuclear bond axis. \label{}}
\end{figure}


\end{document}